\documentclass[aps,prd,twocolumn,superscriptaddress,showpacs,preprintnumbers,amsmath,amssymb]{revtex4-1}
\usepackage{graphicx}
\usepackage{dcolumn}
\usepackage{bm}
\graphicspath{{ps/}}

\usepackage{amsfonts,amssymb,euscript}
\usepackage{amsmath}
\usepackage[T1]{fontenc}
\usepackage{cases}
\usepackage{float}
\usepackage{color}
\usepackage{slashed}
\usepackage{array,makecell}
\usepackage{boldline,multirow}

\usepackage[usenames,dvipsnames,svgnames,table]{xcolor}

\usepackage[colorlinks=true]{hyperref}

\begin{document}

\title{\boldmath 
Horizon hair of extremal black holes and measurements at null infinity}
\affiliation{DPMMS, Cambridge University, Cambridge, CB3 0WB, UK}
\affiliation{Department of Mathematics, UCLA, Los Angeles, CA 90095, USA}
\affiliation{Department of Mathematics, University of Toronto, Toronto, M1C 1A4, Canada}

 \author{Y.~Angelopoulos}\affiliation{Department of Mathematics, UCLA, Los Angeles, CA 90095, USA} 
  \author{S.~Aretakis}\affiliation{Department of Mathematics, University of Toronto, Toronto, M1C 1A4, Canada} 
  \author{D.~Gajic}\affiliation{DPMMS, Cambridge University, Cambridge, CB3 0WB, UK} 

\begin{abstract}

It is shown that the conserved charges on the event horizon and the Cauchy horizon associated to scalar perturbations on extremal black holes are externally measurable from null infinity. This suggests that these charges have the potential to serve as an observational signature.  The proof of this result is based on obtaining \textit{precise} late-time asymptotics for the radiation field of outgoing perturbations.

\end{abstract}

\maketitle

\section{Introduction}
\label{sec:Introduction}

Extremal black holes play a fundamental role in general relativity, high energy physics and astronomy. It has been reported  \cite{rees2005} that $70\%$ of stellar black holes (such as Cygnus X-1 \cite{extremalobservation2} and GRS 1915+105 \cite{extremalobservation1}) are near-extremal, suggesting that near-extremal black holes are ubiquitous in the universe. It has also been argued \cite{brenneman-spin} that many supermassive black holes (such as the ones in the center of MCG--06-30-15 \cite{xrayextremal} and NGC 3783 \cite{brennemanpaper}) are near-extremal. The spins of the astrophysical black holes in all these works are below the widely predicted upper bound $a \approx 0.998M$, which is called the Thorne limit \cite{Thorne1974}. Note that more recent works suggest that it may be possible to go beyond the Thorne limit in the astrophysical setting \cite{Skadowski2011}. Identifying observational \emph{signatures} that indicate the presence of black holes that are sufficiently close to extremality may be fruitful for investigating whether astrophysical black holes with spins beyond the Thorne limit exist; see for example \cite{gralla2016}. Extremal black holes also have interesting theoretical properties. For example, they saturate geometric inequalities for the total mass, angular momentum and charge \cite{dainprl6, reiris2013, alaeeprl}. Moreover, they have zero temperature and hence they play an important role in the study of Hawking radiation \cite{haw95} and in string theory \cite{stromingerextremalentropy}. Their near-horizon limits yield new solutions to the Einstein equations with conformally invariant properties classified in \cite{nearhorizonrn, h07, hol10}. Applications in quantum gravity have been obtained in \cite{strominger-extremal-holography, conformalnhek, cft-extremal-cosmo} and gravitational and electromagnetic signatures of the near-horizon geometry have been presented in \cite{gralla2016, grallastrominger}. 

An important aspect of extremal black holes is that they exhibit intriguing dynamical properties. Perturbations of various types suffer from a ``horizon instability'' \cite{aretakis4,aretakis1,aretakis2,hj2012,hm2012} according to which derivatives transversal to the event horizon of dynamical quantities grow asymptotically in time along the event horizon. The source of this instability is the existence of a charge (i.e.\ a surface integral) $H$  which is conserved along the horizon. We remark that, under the presence of superradiance,  a sequence of  zero-damped quasinormal modes has been found \cite{glampedakisfull,zeni13} leading to an amplified version of the horizon instability \cite{zimmerman1} on such backgrounds. For another type of gravitational instability, we refer to \cite{luis}. 

In this letter, we address the no-hair hypothesis in the case of extremal black holes. The no-hair hypothesis postulates that the only externally observable classical parameters of black hole spacetimes are the mass, electric charge and angular momentum; all other dynamical information (known as ``hair'') is ``lost'' behind the event horizon rendering it permanently inaccessible to external observers. The uniqueness theorems (see e.g.~\cite{alexakisduke}) and stability theorems (see e.g.~\cite{Dafermos2016}) provide a first confirmation of the no-hair hypothesis for sub-extremal black holes. In the extremal case, however, the aforementioned conserved charge $H$ on the event horizon may be viewed as another classical parameter of the black hole.
On the other hand, all natural quantities (e.g.~translation-invariant derivatives of all orders) decay in time away from the horizon.  For this reason, $H$ can be thought of as ``horizon hair'' for the extremal black hole \cite{harvey2013}. 

An open problem discussed in \cite{hm2012, ind2018} is the measurement of the horizon hair $H$ by \textbf{far-away} observers who receive radiation from the near-horizon region. Such observers live in the spacetime region where the distance $r$ from the black hole is large and comparable in size to $t$, the standard time coordinate. This region is modelled by null infinity. In this letter, \textit{we show that the horizon hair $H$ of scalar perturbations on Extremal Reissner--Nordstr\"{o}m (ERN) is measurable from null infinity}, providing thus a resolution to the above open problem (see Section \ref{sec:ObservingTheHorizonInstabilityFromNullInfinity}). This result has not been seen before in the literature and appears here for the first time. Previous works \cite{hm2012, ori2013} (see Section \ref{sec:AsymptoticsForERN} for a review and more details) showed that the horizon hair can be read off at constant distances $r$ or distances $r$ that are much smaller than $t$, but they did not address the measurement of $H$ from null infinity.

  Our result suggests that 1) \textit{extremal black holes admit classical externally measurable hair} and 2) \textit{the horizon instability could potentially serve as an observational signature}. 
 Another implication  is that scalar perturbations also admit a conserved charge inside the black hole, on the Cauchy (inner) horizon, whose value is equal to that of the event horizon hair $H$. This directly implies that the conserved charge on the Cauchy horizon is also measurable from null infinity. Hence, \textit{our result provides a new mechanism that can be used to read off information at the event horizon and at the Cauchy horizon from null infinity}. We further note that our mathematically rigorous argument uncovers a new connection with soft hair (see also the discussion in Section \ref{sec:ReviewOfSubExtremalRN}).

\section{The horizon hair $H[\psi]$ of ERN}
\label{sec:TheHorizonHairHPsi}

We next briefly recall the horizon instability of extremal black holes.  We will consider scalar perturbations $\psi$ solving the wave equation $\Box_g\psi=0$ where $g$ is the ERN metric which in ingoing EF coordinates $(v,r,\theta,\varphi)$ takes the form
\[g=-Ddv^2+2dvdr+r^2(d\theta^2+\sin^2\theta d\varphi^2),\]
where $D=\left(1-\frac{M}{r}\right)^2$. The event horizon corresponds to $\mathcal{H}^{+}=\left\{r=M\right\}$. The vector field $T=\partial_v$ is stationary and normal to $\mathcal{H}^{+}$, whereas $\partial_r$ is translation-invariant $([\partial_r ,T]=0)$ and transversal to $\mathcal{H}^{+}$. Let  $\Sigma_0$ be a spherically symmetric Cauchy hypersurface which crosses the event horizon and terminates at null infinity (for example, we can take $\Sigma_0$ to be $\{v=0 \}$ for $r\leq 2M$ and $\{u=0 \}$ for $r\geq 2M$, where $u,v$ are the standard double null coordinates) and let $\Sigma_{\tau}=F_{\tau}(\Sigma_0)$ where $F_{\tau}$ is the flow of the vector field $T$. We denote by $\partial_{\rho}$ the radial vector field that is tangential to $\Sigma_{\tau}$ and satisfies $\partial_{\rho}r=1$. Let $S_{\tau}=\mathcal{H}^{+}\cap \Sigma_{\tau}$. Then, the following surface integrals
	\begin{equation}
H[\psi]:=-\frac{M^2}{4\pi}\int_{S_{\tau}}\partial_r (r\psi) \, d\Omega
\label{introhorizonH}
\end{equation} 
	are independent of $\tau$ and hence are conserved on $\mathcal{H}^{+}$ for all solutions $\psi$ to the wave equation on ERN. Here $d\Omega=\sin\theta d\theta d\varphi$.  We will refer to $H[\psi]$ as the horizon hair of $\psi$. In fact, there exists an infinite number of analogous conserved charges $H_{\ell}[\psi]$ for each angular momentum $\ell$ appearing in the spherical harmonic decomposition of $\psi$ \cite{aretakis2}, with $H[\psi]=H_0[\psi]$.
	
	We next consider \textit{outgoing} perturbations which arise from compactly supported and horizon penetrating ($H\neq 0$) initial data. It turns out that the following instability results on $\mathcal{H}^{+}$ \cite{aretakis1,aretakis2}: 1) \underline{\textbf{Non-decay}}: $\partial_r \psi|_{\mathcal{H}^{+}}\sim -\frac{1}{M} H[\psi] $ as $\tau\rightarrow \infty$, 2) \underline{\textbf{Blow-up}}: $\partial_r \partial_r \psi|_{\mathcal{H}^{+}}   \sim \frac{1}{M^3}H[\psi] \cdot \tau$  	 as $\tau\rightarrow \infty$. 
More generally $\partial_r^{k}\psi|_{\mathcal{H}^{+}}\sim c_k\cdot H[\psi]\cdot \tau^{k-1}$ where 
$ c_k  =  (-1)^{k}\frac{1}{M^3}\frac{k!}{(2M^2)^{k-1}}$ for $k\geq 1$.  The quantity $H$ can be given a physical interpretation by considering the energy density measured by incoming observers at $\mathcal{H}^+$: $\boldsymbol{T}_{rr}[\psi]\sim M^{-6}\cdot H^2[\psi]$, where $\boldsymbol{T}$ is the energy-momentum tensor, and hence does not decay along $\mathcal{H}^{+}$. On the other hand, all physically relevant quantities decay in time \underline{away} from the horizon. Murata--Reall--Tanahashi's numerical simulations \cite{harvey2013} of the evolution of the Einstein--Maxwell-scalar field system for perturbations of ERN suggest that the horizon instability persists in the fully non-linear setting. This instability is also relevant for \emph{near-extremal} black holes where it is expected to be a \emph{transient phenomenon}, see for example \cite{harvey2013}. For other extensions of this instability we refer to \cite{aretakis4,aretakis1,aretakis2,hj2012,hm2012,sela2, aag1,aretakis2013,harveyeffective,khanna17,aretakis2012,bizon2012,zimmerman1,sela,ind2018,ori2013,murata2012, dd2012,cardoso-2017,zimmerman5,zimmerman2,zimmerman3,berti2017,aretakis3,zimmerman4, gajic}. 

One can also define a conserved charge for scalar perturbations on the Cauchy horizon $\mathcal{C}\mathcal{H}^{+}$ in the black hole interior of ERN (conserved charges can be defined on any hypersurface with vanishing surface gravity \cite{hj2012, aretakisglue}):
	\begin{equation}
\underline{H}[\psi]:=-\frac{M^2}{4\pi}\int_{\underline{S}_{\tau}}\partial_r (r\psi) \, d\Omega,
\label{introhorizonHintro}
\end{equation} 
where $\underline{S}_{\tau}=\{u=\tau\}\cap \mathcal{C}\mathcal{H}^{+}$ and $\partial_r$ is taken with respect to the outgoing EF coordinates $(u,r,\theta,\varphi)$ in the interior region. In contrast to the sub-extremal case, the spherical mean of outgoing perturbations is continuously differentiable at the Cauchy horizon \cite{gajic, harvey2013} and hence $\underline{H}[\psi]$ is well-defined. An important corollary of the precise late-time asymptotics (see Section \ref{sec:RigorousAsymptotics}) is the relation
\begin{equation}
\label{eq:HeqHbar}
H[\psi]=\underline{H}[\psi]
\end{equation}
 for all outgoing perturbations $\psi$.
  \begin{figure}[H]
	\begin{center}
				\includegraphics[scale=0.17]{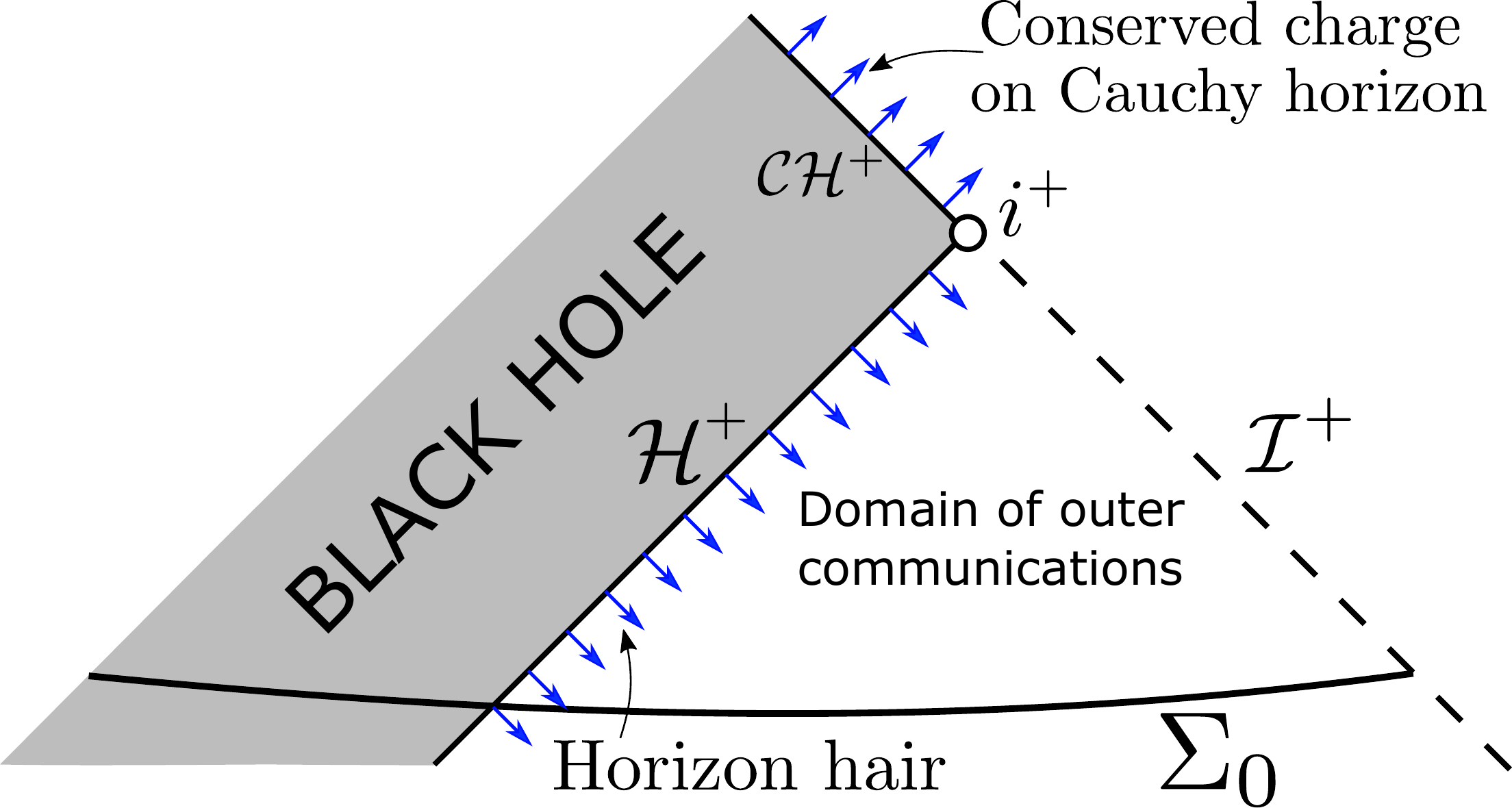}
\end{center}
\vspace{-0.6cm}
\caption{\footnotesize A Penrose diagrammatic representation of the spacetime regions of interest. The conserved charge on the Cauchy horizon is equal to the horizon hair $H[\psi]$ on the event horizon.}\normalsize
	\label{fig:hairinout}
\end{figure}
\vspace{-0.4cm}
\section{Measurements at null infinity}
\label{sec:ObservingTheHorizonInstabilityFromNullInfinity}

We define the following expression involving the radiation field $r\psi|_{\mathcal{I}^{+}}$ of scalar perurbations $\psi$ on any (sub-extremal or extremal) RN spacetime:
\begin{equation}
\boxed{s[\psi]:=\frac{1}{4M}\lim_{\tau\rightarrow \infty}\tau^2\cdot (r\psi)\big|_{\mathcal{I}^{+}} \!\!+\!\frac{1}{8\pi}\int_{\mathcal{I}^{+}\cap\left\{\tau\geq 0\right\}}\!\!\! r\psi\big|_{\mathcal{I}^{+}} \, d\Omega d\tau.}
\label{radfieldexpr}
\end{equation}
In order to compute $s[\psi]$, it actually suffices to know the radiation field for large times $\tau\geq \tau_{\text{late}}$ (for arbitrarily large $\tau_{\text{late}}$). Indeed, the second term on the right hand side of \eqref{maineq} is equal to  
\[-\frac{1}{2M}\int_{\mathcal{I}^+\cap\{\tau=\tau_{\text{late}} \}} r^3\partial_{\rho}(r\psi)d\Omega\!+\frac{1}{8\pi}\int_{\mathcal{I}^{+}\cap\{\tau\geq \tau_{\text{late}}\}}\!\!\! r\psi d\Omega d\tau.\]
We obtain the following identity on sub-extremal and extremal RN:
\begin{numcases}{s[\psi]=}
\label{maineq}
H[\psi] & in extremal RN,\\
\label{maineqsub}
0 & in sub-extremal RN,
\end{numcases}
where in \eqref {maineq} $\psi$ is an outgoing scalar perturbation on ERN and in \eqref{maineqsub} $\psi$ is an initially compactly supported scalar perturbation on sub-extremal RN. \textbf{Identity \eqref{maineq} appears here for the first time and it shows that the horizon hair $H$ (and consequently, the horizon instability) is measurable purely from null infinity.} A sketch of the derivation of \eqref{maineq} is given in Section \ref{sec:AProofOfTheAsymptotics}. Furthermore, in view of identity \eqref{maineqsub} (discussed further in Section \ref{sec:ReviewOfSubExtremalRN} below) and the fact that $H[\psi] \neq 0$, \textbf{the expression $s[\psi]$ provides an observational signature of extremal black holes}. One could also expect $s[\psi]$ to be useful in a transient sense to provide an observational signature for near-extremal black holes.
The remaining conserved charges $H_{\ell}$ could, in principle, be measured at null infinity in an analogous fashion.
Another consequence of \eqref{maineq}, combined with \eqref{eq:HeqHbar}, is that the conserved charge $\underline{H}$ on the Cauchy horizon can be computed from null infinity.
We further obtain the following identity on hypersurfaces of constant area radius $r=R>M$ in the strong field region:
 \begin{equation}
{H[\psi]=\frac{R-M}{4M}\cdot \lim_{\tau\rightarrow \infty} \tau^{2}\cdot \psi\big|_{r=R}},
 \label{Hfromconstantr}
 \end{equation}
confirming the numerical predictions  of \cite{hm2012} and the heuristic analysis of \cite{sela,ind2018,ori2013}.

\section{Late-time Asymptotics}
\label{sec:RigorousAsymptotics}

\subsection{Review of sub-extremal RN}
\label{sec:ReviewOfSubExtremalRN}

Since higher angular modes  $\psi_{\geq 1}=\psi-\frac{1}{4\pi}\int_{\mathbb{S}^2}\psi\,d\Omega$ decay faster than the spherical mean $\frac{1}{4\pi}\int_{\mathbb{S}^2}\psi\,d\Omega$, it suffices to project to the spherical mean (and hence, equivalently, it suffices to consider spherically symmetric perturbations). For initial data extending to $\mathcal{I}^{+}$ on sub-extremal RN, the unique obstruction to inverting $T$ is the non-vanishing of the Newman--Penrose constant $I[\psi]$, which is a conserved charge along null infinity. This is related to the identity $I[T\bar{\psi}]=0$ for all regular solutions $\bar{\psi}$ to the wave equation. For compactly supported initial data (satisfying $I[\psi]=0$), we can construct the time-integral $\bar{\psi}$ of $\psi$ which satisfies $T\bar{\psi}=\psi$ and has finite Newman--Penrose constant  $I[\bar{\psi}]$. We denote $I^{(1)}[\psi]=I[\bar{\psi}]$. If follows that the unique obstruction to inverting the operator $T^2$ is the non-vanishing of $I^{(1)}[\psi]$. The relevance of $I^{(1)}[\psi]$ became apparent in \cite{paper2} where the precise late-time asymptotics were obtained for compactly supported initial data:
\begingroup
\squeezetable
\begin{table}[H]
    \begin{tabular}{c |c| c } 
\hline
 $\psi|_{\mathcal{H}^{+}}$  &$\psi|_{r=R}$ & $r\psi|_{\mathcal{I}^{+}}$ \\ \hline
 $8I^{(1)}[\psi]\cdot \frac{1}{\tau^{3}}$   &$8I^{(1)}[\psi]\cdot \frac{1}{\tau^{3}}$ & $-2I^{(1)}[\psi]\cdot {\tau^{-2}}-8MI^{(1)}[\psi] \log\tau \cdot \tau^{-3}$ 
\\ \hline
  \end{tabular}
		\vspace{-0.1cm}\caption{\footnotesize Leading order terms in the time asymptotics on sub-extremal RN.}
	\label{subernhorizon}
	\end{table}
\endgroup\normalsize
The following expression of $I^{(1)}[\psi]$ was obtained in terms of compactly supported initial data on $\Sigma_0$ in \cite{paper-bifurcate}:
\begin{equation}
I^{(1)}[\psi]= \frac{M}{4\pi}\int_{\Sigma_{0}\cap\mathcal{H}^{+}}\!\!\psi+\frac{M}{4\pi}\int_{\Sigma_0} \nabla \psi\cdot n_{\Sigma_0},
\label{i1sigma0}
\end{equation}
where the integrals are considered with respect to the induced volume form. It turns out that $I^{(1)}[\psi]$ can be computed from null infinity:
\begin{equation}
I^{(1)}[\psi]=\frac{M}{4\pi}\int_{\mathcal{I}^{+}\cap\left\{\tau\geq 0\right\}} r\psi\big|_{\mathcal{I}^{+}} \, d\Omega d\tau.
\label{i1scri}
\end{equation}
The integral of the radiation field along $\mathcal{I}^{+}$ has appeared before in the work of Luk--Oh \cite{Luk2015} on strong cosmic censorship. 
It is clear from Table \ref{subernhorizon} and identity \eqref{i1scri} that \eqref{maineqsub} holds for perturbations on sub-extremal RN. Note also that the late-time asymptotics along, say, the event horizon depend solely on the integral of the radiation field along null infinity, confirming previous heuristic work predicting dominance of the weak field dynamics in the late-time evolution. 

The  existence of  $I^{(1)}[\psi]$ yields a conservation law which can be recast into an identity between the integral of $r\psi$ along $\mathcal{I}^+$ and an analogous integral along $\mathcal{I}^-$, revealing a tantalizing connection with the presence of a \emph{soft electric hair} \cite{Strominger2014, He2014, Hawking2016, Hawking2017}. Indeed, one may formally derive the null infinity conservation law for $r\psi$ and the conservation of charges associated to soft electric hair for a 2-form $F$ satisfying the Maxwell equations with a source $j$, by integrating the following 4-form equations: $d \star d\psi=0$ and $(d\star F+4\pi\star j)\wedge d\epsilon=0$, respectively, in suitable spacetime regions and applying Stokes' theorem. Here $\epsilon$ denotes an arbitrary smooth function that only depends on the angular coordinates. 

\vspace{-0.5cm}

\subsection{Asymptotics for ERN}
\label{sec:AsymptoticsForERN}

We distinguish three classes of perturbations on ERN: 
\begingroup
\squeezetable
\begin{table}[H]
\begin{center}
  \begin{tabular}{ c|c | c }                                \hline
   Perturbations  & $H$           & $I$                           \\ 
\hline
  outgoing  & $\neq 0$        & $=0$      \\ \hline
static moment &  $\neq 0$       & $\neq 0$             \\ \hline
 ingoing &  $=0$           & $=0$             \\ \hline
  \end{tabular}
  \end{center}
	\vspace{-0.4cm}
 \caption{Types of initial data. Here $H$ denotes the conserved charge on  $\mathcal{H}^{+}$ and  $I$ denotes the Newman--Penrose constant on $\mathcal{I}^{+}$.}
\end{table}
\vspace{-0.4cm}
\endgroup
For outgoing and ingoing perturbations (with compactly supported initial data) we define the constant $I^{(1)}$ as in \eqref{i1sigma0} (or, equivalently, in \eqref{i1scri}).   For ingoing perturbations, we also define 
\begin{equation}
H^{(1)}[\psi]:= \frac{M^2}{4\pi}\int_{\mathcal{H}^{+}}\!\! \psi\big|_{\mathcal{H}^{+}}d\Omega d\tau.
\label{h01}
\end{equation} 
 We refer to $H^{(1)}[\psi]$ as the \textit{time-inverted horizon charge}. 
 A physical interpretation of $H^{(1)}[\psi]$ can be given in terms of the  \textit{dual} scalar field $\widetilde{\psi}$ of $\psi$ defined by $\widetilde{\psi}= \frac{M}{r-M}\psi\circ \Phi\label{dual}
$, where $\Phi$ denotes the Couch--Torrence conformal inversion. It can be easily seen that 1) the duality is self-inverse, 2) $\psi$ solves the wave equation if and only if $\widetilde{\psi}$ solves the wave equation and 3) $H[\psi] =I[\widetilde{\psi}]$.  The latter relation was obtained independently in \cite{bizon2012,hm2012}. It follows that $H^{(1)}[\psi]:= I^{(1)}[\widetilde{\psi}]$. Moreover,  in view of \eqref{i1sigma0} applied to $\widetilde{\psi}$, one may obtain an expression for $H^{(1)}$ in terms of the initial data on $\Sigma_0$. 
We can now present the precise late-time asympotics along the event horizon:
\begingroup
\begin{table}[H]
\begin{center}
    \begin{tabular}{ c|c | c } 
\cline{2-3} &  outgoing data & ingoing data \\ 
\hline
$\psi|_{\mathcal{H}^{+}}$ &   ${2H \cdot \tau^{-1}}$ & $\boldsymbol{-2H^{(1)} \cdot \tau^{-2}}$ 
\\ \hline
$\partial_r \psi|_{\mathcal{H}^{+}}$ &  $ -\frac{1}{M}\cdot H$   & $\boldsymbol{\frac{2}{M^2}\cdot H^{(1)}\cdot {\tau^{-2}}}$ \\ \hline
$\partial_r \partial_r \psi|_{\mathcal{H}^{+}}$	& $ \frac{1}{M^3}\cdot H\cdot\tau$ &  $ \boldsymbol{\frac{1}{M^3}\cdot H^{(1)}}$ \\ \hline
$\partial_r \partial_r \partial_r \psi|_{\mathcal{H}^{+}}$ & $-\frac{3}{2M^5} \cdot H \cdot\tau^2$   & $\boldsymbol{-\frac{3}{M^5}\cdot H^{(1)}\cdot \tau}$  \\ \hline
  \end{tabular}
	\vspace{-0.1cm}\caption{\footnotesize Asymptotics along the event horizon on ERN for outgoing and ingoing perturbations. \textbf{The ingoing asymptotics are new and have not appeared before in the literature.} The  outgoing asymptotics are consistent with \cite{hm2012, sela, ori2013}.  } \normalsize
	\label{ernhorizon}
\end{center}
\end{table}
\vspace{-0.7cm}
\endgroup
 We present below the precise late-time asymptotics away from the horizon:
\begingroup
\squeezetable
\begin{table}[H]
\begin{center}
    \begin{tabular}{ c|c | c } 
	 
  \hline
Data &  $\psi|_{r=R}$ & $r\psi|_{\mathcal{I}^{+}}$ \\ 
\hline
outgoing &   $ \frac{4M}{r-M}H \cdot \tau^{-2}$ & $\!\boldsymbol{\left(4MH-2I^{(1)}\right)\! \cdot \!\tau^{-2}}$
\\ \hline
 static moment & $\!\boldsymbol{4\left(I+\frac{M}{r-M}H \right)\cdot \tau^{-2}}$ & $\boldsymbol{2\cdot I[\psi]\cdot \tau^{-1}}$ \\ \hline
ingoing & $\!\boldsymbol{-8\left( I^{(1)}+\frac{M}{r-M}H^{(1)} \right)\!\cdot\! \tau^{-3}}$ & $\boldsymbol{-2I^{(1)}\cdot \tau^{-2}}$ \\ \hline
  \end{tabular}
\caption{\footnotesize Asymptotics away from the event horizon on ERN and specifically on $r=R>M$ and on null infinity $\mathcal{I}^{+}$. \textbf{The bold terms are new and appear here for the first time. The late-time asymptotics for $r\psi|_{\mathcal{I}^{+}}$, in conjunction with the expression \eqref{i1scri} for $I^{(1)}$, yield \eqref{maineq}. } The asymptotic term for $\psi|_{r=R}$ for outgoing perturbations in the strong field region $\{r=R\}$  is consistent with the results presented in \cite{hm2012, sela, ori2013, Burko2007, zimmerman1,harveyeffective, zimmerman4}.  }
\label{awayerntablerev}
\end{center}
\end{table}
\endgroup \normalsize
 

\section{Sketch of the proof}
\label{sec:AProofOfTheAsymptotics}

In this section we present a summary of the main ideas involved in deriving the late-time asymptotics for \underline{outgoing} spherically symmetric perturbations on ERN. The full details will be presented in the upcoming paper \cite{aag7}. 

\begin{figure}[H]
	\begin{center}
				\includegraphics[scale=0.17]{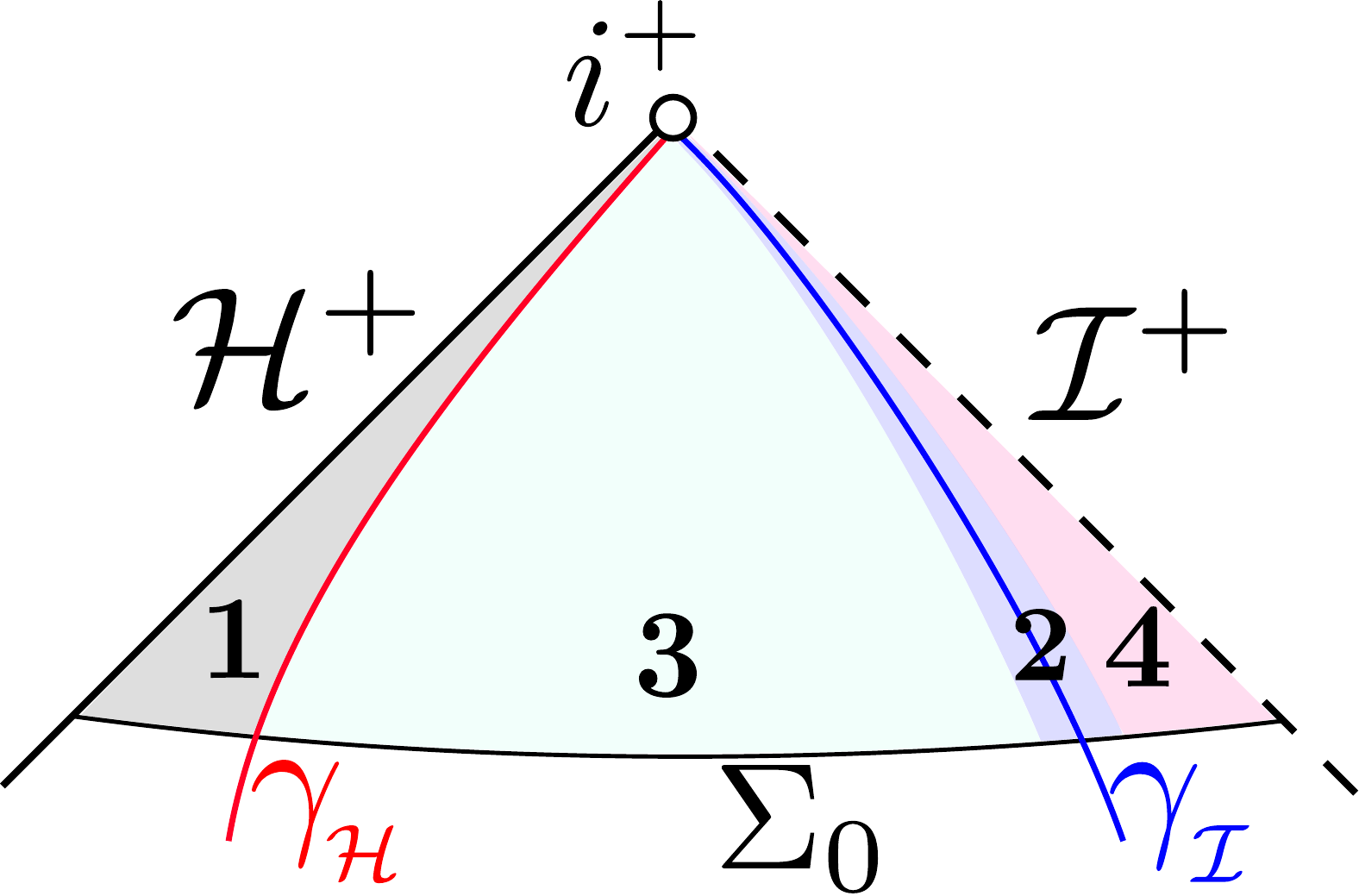}
\end{center}
\vspace{-0.6cm}
\caption{\footnotesize A labeling of the spacetime regions indicating the order in which the late-time asymptotics of $\psi$ are derived. We see that a delicate global study is needed in order to derive the asymptotics on null infinity.  }
	\label{fig:overview}
\end{figure}

\textbf{Step 1.} We obtain the asymptotics for $\psi$ and $T\psi$ on the event horizon and actually in the spacetime region to the left of the hypersurface $\gamma_{\mathcal{H}}=\{r=M+\tau^{\alpha}\}$ for some $3/4<\alpha<1$ (see Figure \ref{fig:overview}). Indeed, we can estimate $\partial_u(r\psi)\sim 2H u^{-2}$ to the left of $\gamma_{\mathcal{H}}$ which after integration from $\gamma_{\mathcal{H}}$ yields asymptotics for $r\psi$ to the left of $\gamma_{\mathcal{H}}$. Here $\partial_u$ is taken with respect to the standard EF double null coordinates $(u,v)$.

\textbf{Step 2.} We derive asymptotics/estimates for the derivative $\partial_{\rho}\psi$ that is tangential to $\Sigma_{\tau}$ as follows:  
integrating the wave equation along $\Sigma_{\tau}$ from the horizon $r=M$ to some  $r>M$ we obtain:\small
\begin{equation*}
\begin{split}
&\ \ \ \ \ Dr^2\partial_{\rho}\psi(r,\tau)=\\ &\boldsymbol{2M^2 T\psi|_{\mathcal{H}^+}(\tau)}
+r^2 T\psi(r,\tau) +\int_{M}^{r}O(r')T\psi+O(r')T^2\psi\,dr'.
\end{split}
\end{equation*}\normalsize The bold horizon term is the leading one: $
\boldsymbol{2M^2 T\psi|_{\mathcal{H}^+}}\sim \boldsymbol{-4MH\cdot \tau^{-2}}$.
We conclude that for any $r>M$:
\begin{equation}
\begin{split}
& \ \ \ \ \ \left|\partial_{\rho}\psi(r,\tau)+\boldsymbol{4MHD^{-1}r^{-2}\tau^{-2}}\right|\\&\leq \:  C\tau^{-\frac{5}{2}+\epsilon}\cdot D^{-\frac{3}{2}}r^{-\frac{1}{2}}+CD^{-1}r^{-2}\tau^{-2-\epsilon}.
\end{split}
\label{partialrhoasymptotics}
\end{equation} 

\textbf{Step 3.} We next obtain the late-time asymptotics for $r\psi$ on $\gamma_{\mathcal{I}}=\{r=\tau^{\alpha}\}$. We use the following splitting identity:
\begin{equation}
\Big. r \psi\Big|_{\gamma_{\mathcal{I}}}\hspace{0.5cm}= \Big.\underbrace{r\partial_\rho (r\psi)\Big|_{\gamma_{\mathcal{I}}}}_{\substack{\text{contribution from} \\ \text{the right side of $\gamma_{\mathcal{I}}$}}}-\Big.\underbrace{r^2 \partial_\rho\psi\Big|_{\gamma_{\mathcal{I}}}}_{\substack{\text{contribution from} \\ \text{the left side of $\gamma_{\mathcal{I}}$}}}.
\label{splittingide}
\end{equation}
We will show that the first (resp.~the second) term  on the right hand side of \eqref{splittingide} can be estimated using properties of the right (resp.~left) side of $\gamma_{\mathcal{I}}$. We introduce a new technique, which we call \textbf{the singular time inversion}. We construct the  time integral $\psi^{(1)}$ of $\psi$ which solves the wave equation  $\square_g\psi^{(1)}=0$ and satisfies $T\psi^{(1)}=\psi$. Since $H[\psi]\neq 0$ we have that  $\psi^{(1)}$ is \emph{singular} at the horizon; in fact, $
(r-M)\cdot \partial_{\rho}\psi^{(1)} =-\frac{2}{M}\cdot H[\psi]$
close to the event horizon. On the other hand, $\psi^{(1)}$  is smooth away from the event horizon and has a well-defined Newman--Penrose constant $I^{(1)}=I[\psi^{(1)}]<\infty$. It can be shown that  $|r\psi^{(1)}|\lesssim \tau^{-1/2+\epsilon}$ as $\tau\rightarrow \infty$ to the right of $\gamma_{\mathcal{I}}$.  The boundedness of $I^{(1)}$ yields $\partial_{\rho}(r\psi^{(1)})|_{\gamma_{\mathcal{I}}}\sim I^{(1)} v^{-2} \sim I^{(1)} \tau^{-2} $  since $v\sim \tau$ and $r\sim \tau^{\alpha}$ along $\gamma_{\mathcal{I}}$. Hence, we obtain $\partial_{\rho}(r\psi)|_{\gamma_{\mathcal{I}}}\sim I^{(1)} \tau^{-3}$ and hence 
$r\partial_{\rho}(r\psi)|_{\gamma_{\mathcal{I}}}\sim r\tau^{-3}\sim \tau^{-3+\alpha}$  along $\gamma_{\mathcal{I}}$. We conclude that this term does \underline{not} contribute to the asymptotics of $r \psi|_{\gamma_{\mathcal{I}}}$. We next derive the precise asymptotics of $r^2 \partial_\rho\psi|_{\gamma_{\mathcal{I}}}$. Integrating the wave equation along $\Sigma_{\tau}$ for $r=R$ to $r=r_{\gamma_{\mathcal{I}}}$ we obtain 
\begin{equation}
\Big|Dr^2\partial_\rho\psi\big|_{{\gamma_{\mathcal{I}}}}-Dr^2\partial_\rho\psi\big|_{r=R}\Big|\lesssim \int^{r_{\gamma_{\mathcal{I}}}}_{R}r|\partial_\rho\big(rT\psi\big)|\,dr.
\label{tarho}
\end{equation}
The right hand side can be shown to be bounded by $\tau^{-2-\epsilon}$ for some $\epsilon>0$ which implies that the asymptotics for $r^2\partial_{\rho}\psi|_{\gamma_{\mathcal{I}}}$ can be derived from the asymptotics of $\partial_{\rho}\psi|_{\{r=R\}}$. We can now apply \eqref{partialrhoasymptotics} for $r=R$ to conclude that the asymptotics for $r^2\partial_{\rho}\psi|_{\gamma_{\mathcal{I}}}$ and $r\psi|_{\gamma_{\mathcal{I}}}$ depend \textbf{only} on $H$. Specifically, we obtain  as $\tau\rightarrow \infty$:
\begin{equation}
r \psi\Big|_{\gamma_{\mathcal{I}}} \sim -{ r^2 \partial_\rho\psi\Big|_{\gamma_{\mathcal{I}}}}  \sim -{Dr^2 \partial_\rho\psi\Big|_{r=R}}\sim 4MH\tau^{-2}.
\label{gammaias}
\end{equation}

\textbf{Step 4.}  Integrating \textbf{backwards} the estimate for $\partial_{\rho}\psi$ of the previous steps from $\gamma_{\mathcal{I}}$  up to $\gamma_{\mathcal{H}}$ and using the asymptotics for $r\psi|_{\gamma_{\mathcal{I}}}$, we obtain the asymptotics for $r\psi$ in the region between $\gamma_{\mathcal{H}}$ and $\gamma_{\mathcal{I}}$.

\textbf{Step 5.} In this last step we derive the asymptotics for $r\psi$ to the right of $\gamma_{\mathcal{I}}$ all the way up to null infinity. We use the construction for the singular time integral $\psi^{(1)}$ once again. Specifically, we  derive the asymptotics of the difference $T(r\psi^{(1)})-T(r\psi^{(1)})|_{\gamma_{\mathcal{I}}}=r\psi-r\psi|_{\gamma_{\mathcal{I}}}$  in terms of $I^{(1)}=I[\psi^{(1)}]$:
\begin{equation*}
\Big|r\psi|_{\mathcal{I}^{+}}(\tau)-r\psi|_{\gamma_{\mathcal{I}}}(\tau)+ 2I^{(1)}\cdot \tau^{-2}\Big|\lesssim C \tau^{-2-\epsilon}.
\end{equation*}
Plugging in the asymptotics \eqref{gammaias} of $r\psi|_{\gamma_{\mathcal{I}}}$ yields the asymptotics of the radiation field $r\psi$  as in Table \ref{awayerntablerev}.

\vspace{-0.15cm}
\vspace{-0.4cm}\normalsize

\section{Concluding remarks}
\label{sec:ConcludingRemarks}

The physical relevance of our results stems from the expectation that the horizon hair of \textit{axisymmetric} scalar perturbations on Extremal Kerr (EK) can be analogously measured from null infinity. Even though the late-time behavior for fixed non-zero azimuthal modes on EK has been derived by Casals--Gralla--Zimmerman \cite{zimmerman1}, the precise late-time asymptotics are not known. In fact, a very exciting problem would be to examine potential contributions of the near-horizon geometry to the precise late-time asymptotics for \textbf{general} (without any symmetry assumptions) scalar, electromagnetic and gravitational  perturbations on EK.  A closely related problem is to probe the measurability properties of the Lucietti--Reall gravitational instability \cite{hj2012} of EK from null infinity. The ultimate goal would be of course to study the fully non-linear perturbations of EK in the context of the Einstein-vacuum equations.  A simplified but still very interesting problem would be to obtain analogous measurability results for the Murata--Reall--Tanahashi spacetimes \cite{harvey2013}.

\section{Acknowledgements}
\label{sec:Acknowledgements}

We thank Harvey Reall for his insightful comments. S. Aretakis acknowledges support through NSF grant DMS-1265538, NSERC grant 502581, an Alfred P. Sloan Fellowship and the Connaught Fellowship 503071.


\end{document}